\documentclass[aip,reprint]{revtex4-1}
\usepackage{graphicx}

\begin{document}


\title{High-charge divergent electron beam generation from high-intensity laser interaction with a gas-cluster target} 



\author{P. Koester}
\affiliation{CNR-INO Istituto Nazionale Ottica, Via Moruzzi 1, 56124 Pisa, Italy}
\author{G.C. Bussolino}
\affiliation{CNR-INO Istituto Nazionale Ottica, Via Moruzzi 1, 56124 Pisa, Italy}
\author{G. Cristoforetti}
\affiliation{CNR-INO Istituto Nazionale Ottica, Via Moruzzi 1, 56124 Pisa, Italy}
\author{A. Faenov}
\affiliation{Joint Institute for High Temperatures, Russian Academy of Science (RAS), 13-2, Izhorskaya st., Moscow, 125412, Russia}
\author{A. Giulietti}
\affiliation{CNR-INO Istituto Nazionale Ottica, Via Moruzzi 1, 56124 Pisa, Italy}
\author{D. Giulietti}
\affiliation{Dipartimento di Fisica, Universit\`a di Pisa, Largo Pontecorvo 3, 56127 Pisa, Italy}
\affiliation{INFN Sezione di Pisa, Largo Pontecorvo 3, 56127 Pisa, Italy}
\author{L. Labate}
\affiliation{CNR-INO Istituto Nazionale Ottica, Via Moruzzi 1, 56124 Pisa, Italy}
\affiliation{INFN Sezione di Pisa, Largo Pontecorvo 3, 56127 Pisa, Italy}
\author{T. Levato}
\affiliation{Universit\`a di Roma ``Tor Vergata'', Dip. Ingegneria Industriale, Via del Politecnico 1, 00133 Roma, Italy}
\affiliation{CNR-INO Istituto Nazionale Ottica, Via Moruzzi 1, 56124 Pisa, Italy}
\affiliation{INFN Sezione di Pisa, Largo Pontecorvo 3, 56127 Pisa, Italy}
\author{T. Pikuz}
\affiliation{Joint Institute for High Temperatures, Russian Academy of Science (RAS), 13-2, Izhorskaya st., Moscow, 125412, Russia}
\author{L.A. Gizzi}
\affiliation{CNR-INO Istituto Nazionale Ottica, Via Moruzzi 1, 56124 Pisa, Italy}
\affiliation{INFN Sezione di Pisa, Largo Pontecorvo 3, 56127 Pisa, Italy}


\date{\today}

\begin{abstract}
We report on an experimental study on the interaction of a high-contrast 40\,fs duration 2.5\,TW laser pulse with an argon cluster target. A high-charge, homogeneous, large divergence electron beam with moderate kinetic energy ($\approx2$\,MeV) is observed in the forward direction. The results show, that an electron beam with a charge as high as 10\,nC can be obtained using a table-top laser system. The accelerated electron beam is suitable for a variety of applications such as radiography of thin samples with a spatial resolution better than 100\,$\mu$m. 
\end{abstract}


\maketitle 


The generation of electron beams through laser driven electron acceleration in an underdense plasma\,\cite{Tajima_PRL79} is a promising approach for next generation electron accelerators. The interaction of a high-intensity ultra-short laser pulse with a gasjet target was studied extensively in the past years\,\cite{Malka_SC02, Giulietti_SSCP10}. Recently a new type of target, the cluster targets, has received much attention due to its unique properties\,\cite{Kishimoto_HFS01}. Gas-cluster targets are characterized by a relatively low average density with localized regions of solid density. Efficient laser pulse propagation and enhanced laser energy absorption were observed in clusterized gases for non-relativistic laser intensities. Electron acceleration in cluster targets to energies of several tens to hundreds of MeV was demonstrated experimentally for laser intensities above $10^{19}$\,W/cm$^2$\,\cite{Fukuda_PLA07, Zhang_APL12}. Clustered gas presents some advantages with respect to the usual gas targets. In fact, the increased laser absorption allows higher values of ionization degree in relatively high density plasmas (10$^{19}$-10$^{20}$\,e/cm$^3$). The parameters of the plasmas so produced are particularly favourable  for the acceleration of high charge and energetic electron bunches. As a matter of fact the maximum accelerating electric field, related to the plasma waves in which the acceleration process develops, scales as the square root of the plasma density and the charge of the accelerated electron bunches increases with the density. In fact, the typical charge of electron bunches produced in gas-jet targets is of the order of a few tens to a few hundreds of pC, much lower than the results reported here. So the clustered gas jets are considered very promising targets for innovative sources of high charge bunches of energetic electrons. 

Such sources could have several applications. Among them the development of compact electron injectors for conventional accelerators, characterized by a high charge (nC) and with the supplementary advantage of an easy synchronization with other apparatus (e.g. a laser) for pump and probe experiments. Another immediate application is the direct utilization of the bunches of the accelerated electrons for electron radiography. In the frame of the Charged Particle Radiography\,\cite{Merrill_NIMB07, Faenov_APL09}, electron contact radiography is one of the possible approaches to the development of imaging techniques with high spatial resolution for thin objects, which is important for a variety of applications including imaging of biological samples. It requires an electron beam with relatively low energy (a few MeV), such that the penetration depth of the electrons is of the order of the thickness of the sample. The high divergence of the accelerated electrons that characterize laser-plasma acceleration\,(LPA) in some experimental conditions is suitable for such application, allowing to irradiate larger samples without the use of electron optics. Moreover the reduced energy spread is not a ``must'' as for the applications in the new acceleration techniques. Homogeneity and high charge are additional important characteristics of the electron beam to be suitable for electron radiography.

Here we report on the experimental results obtained from the interaction of a 2.5\,TW laser pulse with an Ar cluster-gas target. A high-charge, divergent electron beam of modest energy (up to a few MeV) is observed in the forward direction.

The experiment was performed at the Intense Laser Irradiation Laboratory in Pisa.
The laser system delivers pulses with an energy up to 100\,mJ on target. In the described experiment the energy on target is 80\,mJ/pulse. The laser beam (800\,nm, 40\,fs) and is focused by means of an f/5 off-axis parabola into the gasjet. The nominal peak laser intensity is $1.7\times10^{18}$\,W/cm$^2$ with a contrast of $5\times10^8$ on a nanosecond time scale. The supersonic gasjet nozzle is rectangular with dimensions $4\times1.2$\,mm$^2$ and the laser propagation direction is parallel to the shorter edge of the nozzle. The laser is focused close to the entrance border of the gasjet at a vertical distance of 0.5\,mm from the aperture. The gas used in the experiment is Ar at a backing pressure between 45-50\,bar. During the expansion of the Ar gas in the nozzle, clusterization of the gas occurs as discussed below. A Kodak Lanex Regular scintillating screen is mounted at a distance of about 15\,cm from the gasjet nozzle for the characterization of the spatial profile of the electron beam. An imaging system is used to view the Lanex screen from outside the vacuum chamber. A magnetic spectrometer with an entrance slit of 0.5\,mm width is inserted between the nozzle and the Lanex screen at a distance of 8\,mm from the Lanex screen for the characterization of the energy distribution of the accelerated electron beam. The charge of the electron beam is measured by means of a dosimetric film (Gafchromic MD-55). A 15\,$\mu$m thick Al filter is placed in front of the dosimetric film to avoid direct irradiation by the transmitted laser light and plasma self emission.

\begin{figure}
\includegraphics[width=8cm]{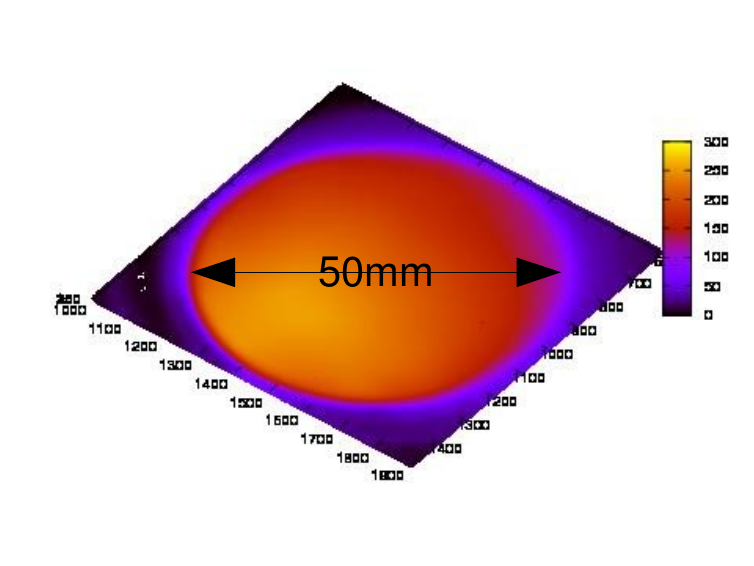}%
\caption{Typical spatial distribution of the accelerated electron beam as detected by the scintillating screen at a distance of 15\,cm from the gasjet nozzle. The electron beam was obtained from the interaction of a 40\,fs laser pulse at an intensity of $2\times10^{18}$\,W/cm$^2$ with an Ar cluster target (backing pressure 50\,bar). \label{FigLanex}}%
\end{figure}

\begin{figure}
\includegraphics[width=8cm]{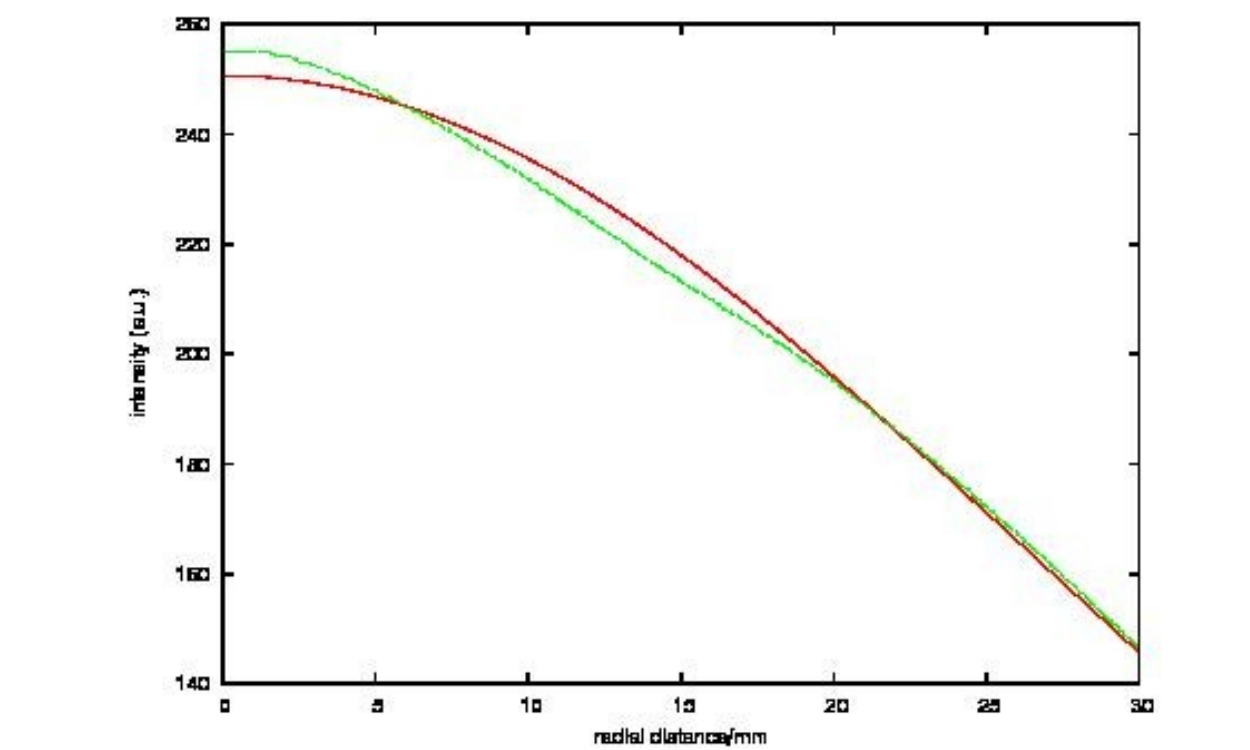}%
\caption{Radial lineout of the spatial distribution in fig.\ref{FigLanex} and gaussian fit.\label{FigFit}}%
\end{figure}

The spatial distribution of the electron beam detected by means of the scintillating screen as obtained from a single shot of the driving laser is displayed in fig.\ref{FigLanex}. 
The electron beam shows a very homogeneous spatial distribution over a wide region. In this case, the electron beam is decentered on the Lanex screen in order to get information on the outer, low intensity regions. Over the whole diameter of the scintillating screen of 5\,cm the signal changes by at most a factor of two. A lineout along the diameter of the graph in fig.\ref{FigLanex} is shown in fig.\ref{FigFit} together with a gaussian fit $f(x)=A\exp^{\frac{(x-x_c)^2}{2\sigma^2}}$. The lineout of the data shows a smooth spatial profile and the width of the gaussian fit results $\sigma=26.8$\,mm. Taking into account the distance from the nozzle to the scintillating screen the electron beam divergence is calculated to be 0.40\,rad at FWHM.

\begin{figure}
\includegraphics[width=8cm]{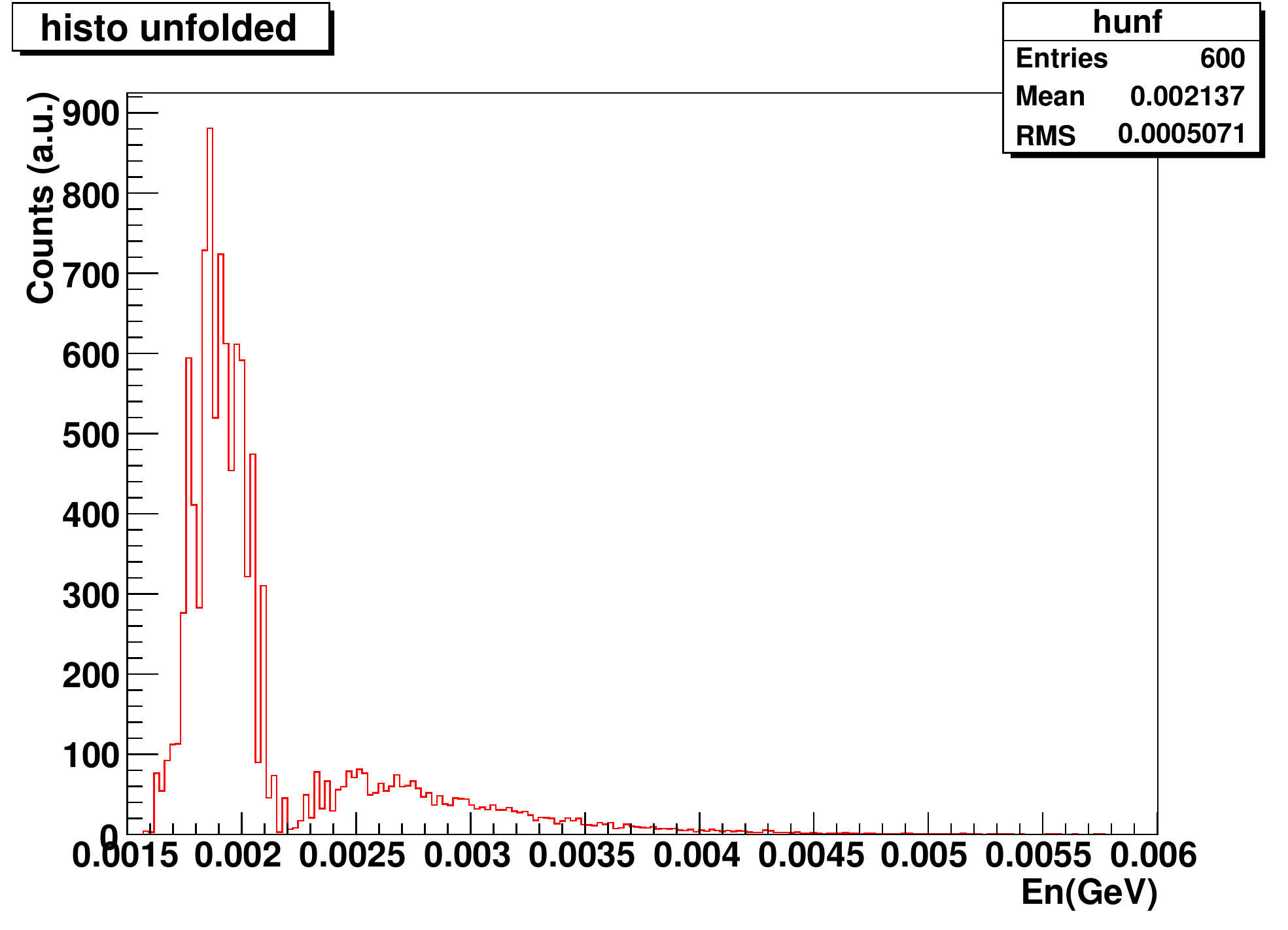}%
\caption{Typical energy distribution of the electron beam obtained from the interaction of a 40\,fs laser pulse at an intensity of $2\times10^{18}$\,W/cm$^2$ with an Ar cluster target (backing pressure 46\,bar).\label{FigSpectrum}}%
\end{figure}

The energy distribution of the electron beam above the detection limit of 1.5\,MeV is shown in fig.\ref{FigSpectrum}. The spectrum was obtained from the interaction of the laser pulse with the Ar cluster target at a backing pressure of 46\,bar. The spectrum shows a bright peak at about 1.9\,MeV with a low intensity tail at higher energies.

\begin{figure}
\includegraphics[width=8cm]{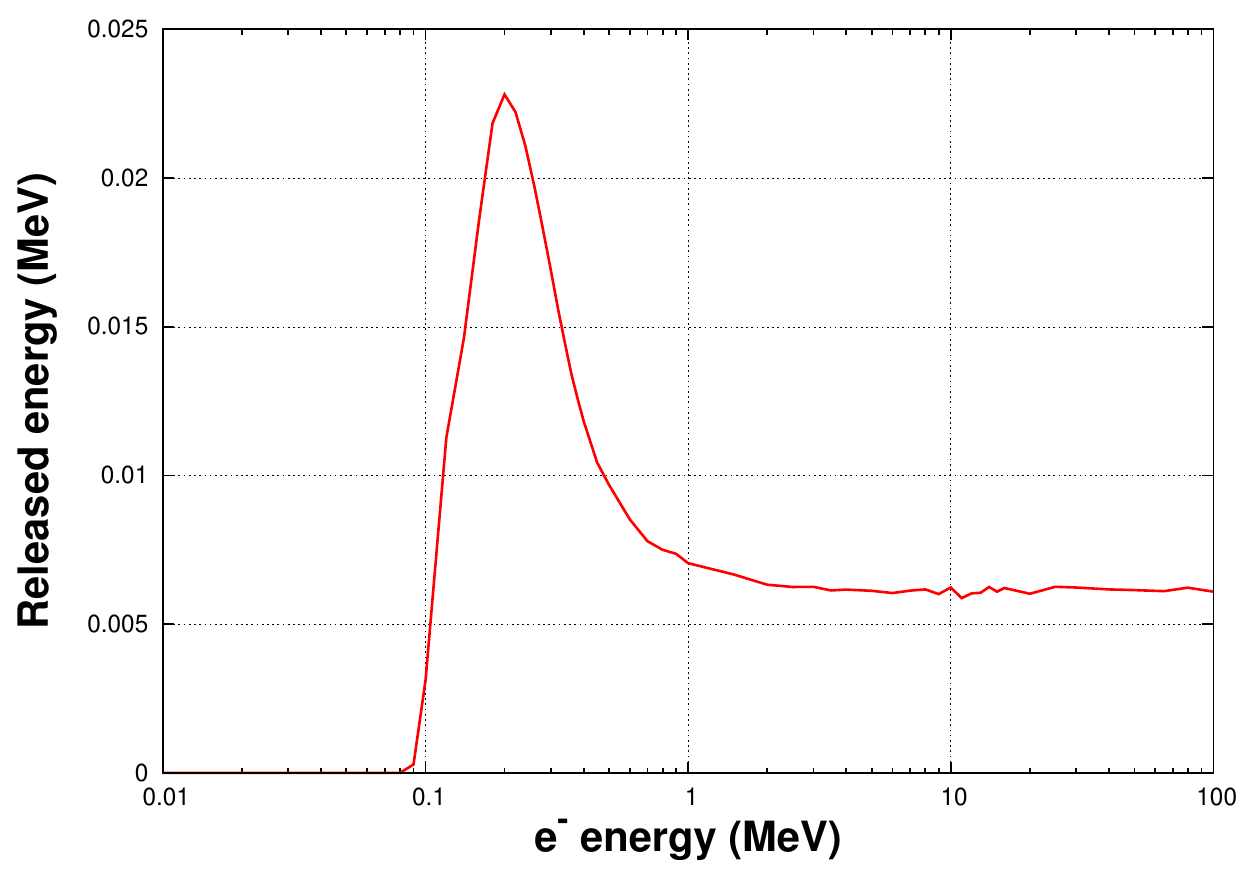}%
\caption{Energy released in the active layers of a GAFCHROMIC MD-55 film as a function of incoming electron kinetic energy. The resutls are obtained from Montecarlo simulations using Geant4 libraries for the experimental setup including an Al filter of 15\,$\mu$m thickness. \label{FigMonteCarlo}}%
\end{figure}

Information on the charge of the electron beam was retrieved through the analysis of the signal detected on a dosimetric film for a series of 36 pulses of the driving laser. 
From the sensitometric response of Gafchromic MD-55 densitometry film\,\cite{Chair_MP98} 
and the measured optical density of the exposed film (OD=0.77) we get a peak dose of about 25 Gy in the center of the spatial distribution of accelerated  
electrons. Considering the thickness (16\,$\mu$m) of the two active layers  
and their density (1.08\,g/cm$^3$) we can obtain the surface mass density  
and finally the energy released by the energetic electrons on the  
dosimetry film (8.64$\times$10$^{-5}$\,J/cm$^2$). A Monte Carlo simulation on the energy released by energetic electrons on MD-55 dosimetry film is  
reported in Figure\,\ref{FigMonteCarlo}. 
As we can see 200\,keV electrons release 22\,keV, while electrons with energy of the order or greater than 1\,MeV release 6\,keV. Thus the charge of the electron beam is in the range 5\,nC to 18\,nC, where the lower limit applies to a monoenergetic electron beam of 200\,keV kinetic energy and the upper limit occurs for an electron beam with kinetic energy above 1\,MeV. The electron energy distribution below 1\,MeV was not measured during the experiment, as lower energy electrons are trapped inside the magnetic field of the spectrometer and thus are not detected on the Lanex screen. This generates the uncertainty of the charge measurement of 12$\pm$6\,nC.

The generated electron beam was then used to irradiate a sample of a few centimeters size at contact with the radiochromic film detector. A detailed description of the radiographic results can be found elsewhere\,\cite{Bussolino_MPsubmitted}. The spatial resolution was measured from the radiographic image of a sharp edge. Assuming a gaussian point-spread-function, the RMS spotsize was found to be 60\,$\mu$m.

\begin{figure}
\includegraphics[width=8cm]{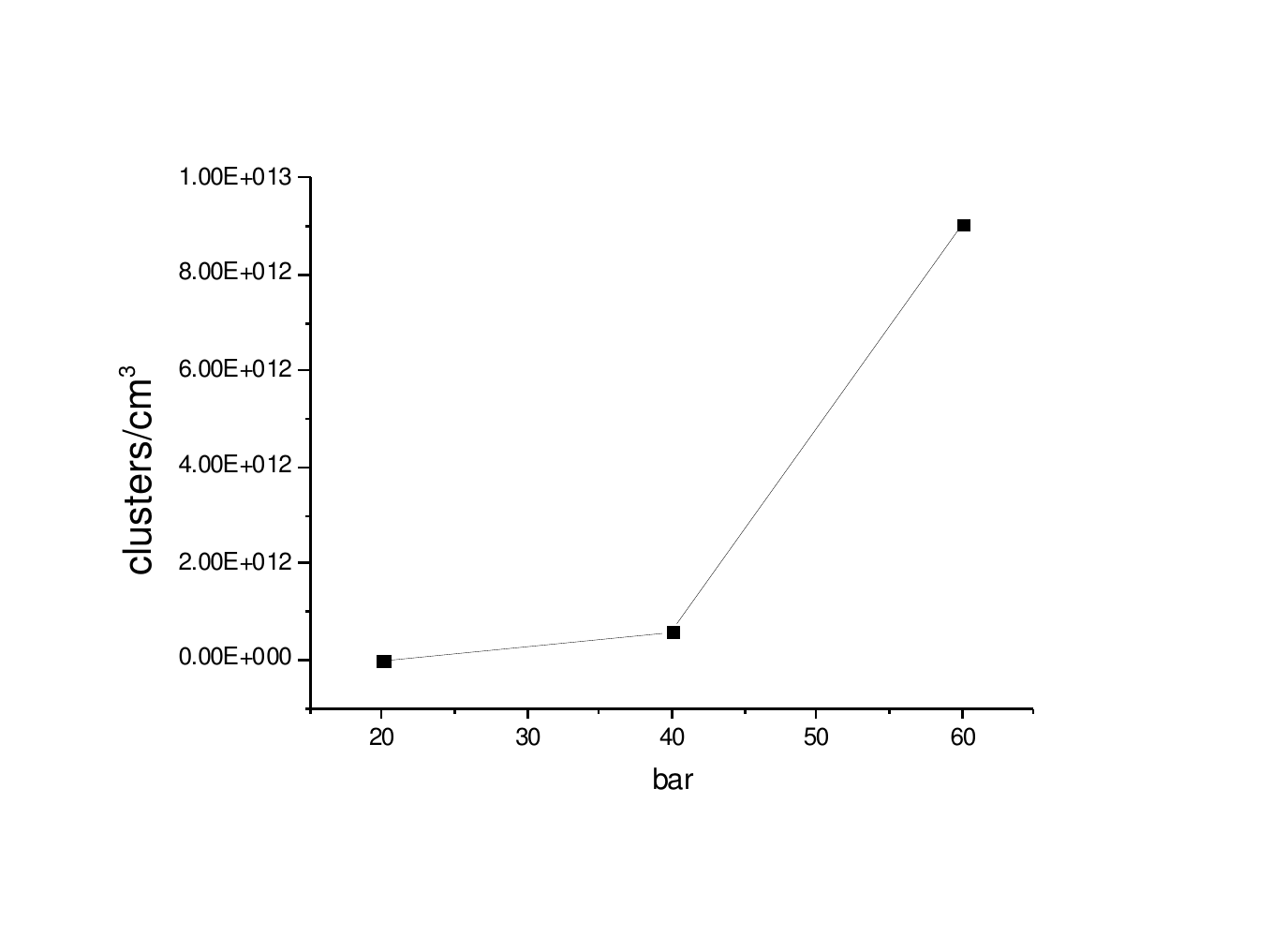}%
\caption{Cluster density as a function of Ar gas backing pressure.\label{Fig:Clusters}}%
\end{figure}

The characterization of the electron beam shows that a high charge, homogeneous and divergent electron beam with moderate kinetic energy of ~2\,MeV was generated through the interaction of the laser pulse with a supersonic Ar gasjet target. It is well known that clusterization of the Ar gas occurs in our experimental conditions. In order to get information on the cluster parameters, simulations were performed in a simpler conical geometry. The simulations show that Ar clusters with a radius between 40\,nm (at 40\,bar backing pressure) and 20\,nm (at 60\,bar) are generated. The concentration of clusters increases rapidly for backing pressures between 40 and 60\,bar\,(see Fig.\,\ref{Fig:Clusters}). From an experimental point of view, it is observed that both the stability and the charge of the electron beam increases with increasing backing pressure, indicating that the presence of clusters is crucial for the generation of the high charge electron beam. In fact, the efficient absorption of lasers in clusterized gases leads to high ionization stages of the Ar atoms and thus to a high electron density\,\cite{Kishimoto_HFS01}. Ionization stages of 16 were reached in Ar cluster targets for laser intensities similar to the ones used in the experiment\,\cite{PhysRevA.63.032706}.  Thus, for a neutral gas density of 4$\times$10$^{19}$\,cm$^{-3}$ as expected for the used backing pressure, plasma electron densities of 10$^{21}$\,cm$^{-3}$are reached. This might explain the huge electron beam charge observed in the experiment. On the other hand, the high plasma density does not lead necessarily to a higher kinetic energy of the accelerated electrons. The accelerating plasma wave is distorted due to the inhomogeneous plasma density generated by the interaction of the laser pulse with the clusterized gas. Therefore, the acceleration meccanism might be less efficient and generate electrons with modest kinetic energy.

In conclusion, high-charge multi-MeV electron bunches generated in LPA experiments using clustered gas targets show suitable characteristics for several applications, among them the development of compact injectors of electrons for conventional accelerators and innovative source for pulsed electron radiography. The results presented indicate that stable and under control sources can be set-up using a multi-TW laser and supersonic gas jet.


%
%

%


\bibliography{BibKoester}

\end{document}